\documentclass[aps,prl,twocolumn,groupedaddress]{revtex4}

\usepackage{graphicx} 

\begin{document}

\renewcommand{\thefootnote}{\alph{footnote}}

\preprint{APL, Y. Takahashi}

\title{
Strong photo-absorption by a single quantum wire in waveguide-transmission spectroscopy 
}

\author{
Yasushi Takahashi\footnote[1]{Electronic mail: taka8484@issp.u-tokyo.ac.jp}, 
Yuhei Hayamizu, Hirotake Itoh, 
Masahiro Yoshita\footnote[2]{Also a visiting scientist at Bell laboratories, Lucent Technologies}, 
and Hidefumi Akiyama$^\dagger$}

\affiliation{
Institute for Solid State Physics, University of Tokyo, and CREST, JST,\\
5-1-5 Kashiwanoha, Kashiwa, Chiba 277-8581, Japan
}

\author{Loren N. Pfeiffer and Ken W. West}

\affiliation{Bell Laboratories, Lucent Technologies, 600 Mountain Avenue, Murray Hill, NJ 07974}

%
%
\date{Feb. 15, 2005}

\begin{abstract}

We measured the absorption spectrum of a single T-shaped, 14$\times$6 nm lateral-sized quantum wire embedded in an optical waveguide using waveguide-transmission spectroscopy at 5 K. In spite of its small volume, the one-dimensional-exciton ground state shows a large absorption coefficient of 80 cm$^{-1}$, or a 98 \% absorption probability for a single pass of the 500-$\mu$m-long waveguide. 
\end{abstract}


\maketitle

\newpage

In quantum wires, interband optical transitions are expected to have enhanced oscillator strength at the lowest energy levels\cite{asada1985}. This enhancement originates from the inverse-square-root divergence of the one-dimensional (1D) joint density of states, and then from the enhanced excitonic effects in 1D systems. In fact, these two effects have been intensively studied in theories\cite{elliott1959,abe1989,ogawa1991,wang2001} and in experiments of photoluminescence (PL) and PL-excitation (PLE) spectroscopy\cite{akiyama2003,itoh2003}.

In optical device applications\cite{arakawa1982,kapon1989,weg1993,yagi2003}, however, small volume of wires, and hence a slight overlap between the quantum wires and the optical waveguide, should reduce modal absorption and gain in devices, which could be a critical drawback. Therefore, it is important to perform direct absorption measurements and characterize the absorption spectra quantitatively for quantum-wire devices. Until now, such data has never been reported.

In this letter, we report the absorption spectrum of a single T-shaped quantum wire (T-wire) at 5 K for light propagating along the T-wire. To measure the absorption spectrum, we employed waveguide-transmission spectroscopy using an appropriate interaction length\cite{weiner1985}. Even though this is a good method for investigating the absorption of small-volume nanostructures, it requires uniformity of the nanostructures over the entire length. Because the T-wire is fabricated using the cleaved-edge overgrowth with molecular-beam epitaxy\cite{pfeiffer1990} and a newly developed  growth-interrupt annealing technique\cite{yoshita2002}, it has unprecedented high uniformity\cite{yoshita2004}, and the absorption-coefficient peaks that correspond to the 1D-exciton ground state, excited state 1D-exciton, and 1D continuum states are clearly resolved and evaluated in the spectrum.

\begin{figure}[bp]
\centerline{\scalebox{1.0}{\includegraphics{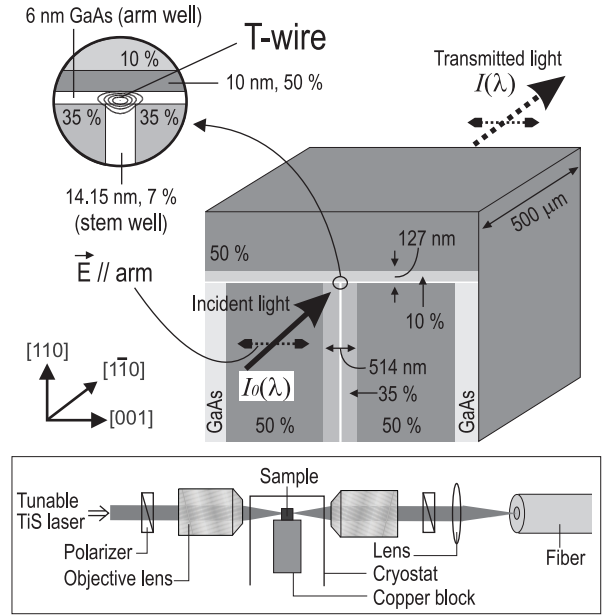}}}
\caption{Schematic views of the single T-wire sample (above) and the waveguide-transmission experiment (below). Percentage for each layer represents the Al content of $\rm{Al_{x}Ga_{1-x}As}$. The sample was attached to a copper block in the cryostat and cooled to 5 K.}
\label{fig1}
\end{figure}

Figure \ref{fig1} shows schematic views of the single T-wire sample and the waveguide-transmission experiment. A single T-wire is formed at the T-shaped intersection of an (001) $\rm{Al_{0.07}Ga_{0.93}As}$ quantum well (stem well) and a (110) GaAs quantum well (arm well). The T-wire size is 14$\times6$ nm. The contour lines show a constant probability for electrons confined in the T-wire ($|\psi|^2=0.2\sim1.0$). The T-wire is embedded in the core of the T-shaped optical waveguide (T-waveguide) that consists of 514-nm-thick (001) layers and 127-nm-thick (110) layers, surrounded by $\rm{Al_{0.5}Ga_{0.5}As}$ cladding layers. A 500-$\mu$m-long Fabry-Perot cavity is formed by the uncoated cleaved facets. Details about the fabrication processes and lasing properties of the single T-wire are reported elsewhere \cite{hayamizu2002}. 

The waveguiding modes are calculated using a finite element method for the T-waveguide model of a 514-nm-thick (001) layer of $\rm{Al_{0.35}Ga_{0.65}As}$ (refractive index $n$=3.38) and a 127-nm-thick (110) layer of $\rm{Al_{0.1}Ga_{0.9}As}$ ($n$=3.56) surrounded by $\rm{Al_{0.5}Ga_{0.5}As}$ layers ($n$=3.28). The refractive indices, $n$ for 1.580 eV at 5 K, used in the calculation are derived from the data for 300 K \cite{aspnes1986} at energy shifted by the band-gap-energy difference for the temperature. The lowest calculated waveguiding mode has a mode refractive index of $n$=3.38, and an optical confinement factor of $\Gamma=4.6\times 10^{-4}$, which is defined as the overlap portion of the 14$\times6$ nm area of the T-wire. This $\Gamma$ value is about two orders of magnitude smaller than that of ordinary multiple-quantum-well lasers.

A tunable continuous-wave titanium-sapphire (TiS) laser was used as the light source and divided into a 1 \% duty ratio. The laser beam was spatially filtered, focused by a 0.5-numerical-aperture objective lens to a spot with a diameter of about 1 $\mu$m on the cavity facet, and then coupled to the T-waveguide. The transmitted light from the other cavity facet was collimated using an objective lens and coupled using another lens to an optical fiber with a core diameter of 50 $\mu$m, in a confocal microscope configuration. This configuration enabled us to eliminate the stray light, which does not propagate through the T-waveguide. Furthermore, we checked the mode pattern of the transmitted light from the T-waveguide using the microscopic imaging method\cite{takahashi2003}, and confirmed a circular pattern that corresponded to the lowest mode for all wavelengths. The transmitted light intensity $I(\lambda)$ from T-waveguide was measured using a liquid-nitrogen-cooled CCD detector with scanning wavelength $\lambda$ of the TiS laser. After removing the sample and adjusting the focus of the lenses, we measured incident light intensity $I_0(\lambda)$ to obtain the transmittance spectrum $T(\lambda) = I(\lambda)/I_0(\lambda)$. Because the T-wire is optically inactive for polarization parallel to the stem well, the polarizations of incident and transmitted light were both set parallel to the arm well using two polarizers. 

\begin{figure}[tbp]
\centerline{\scalebox{1.0}{\includegraphics{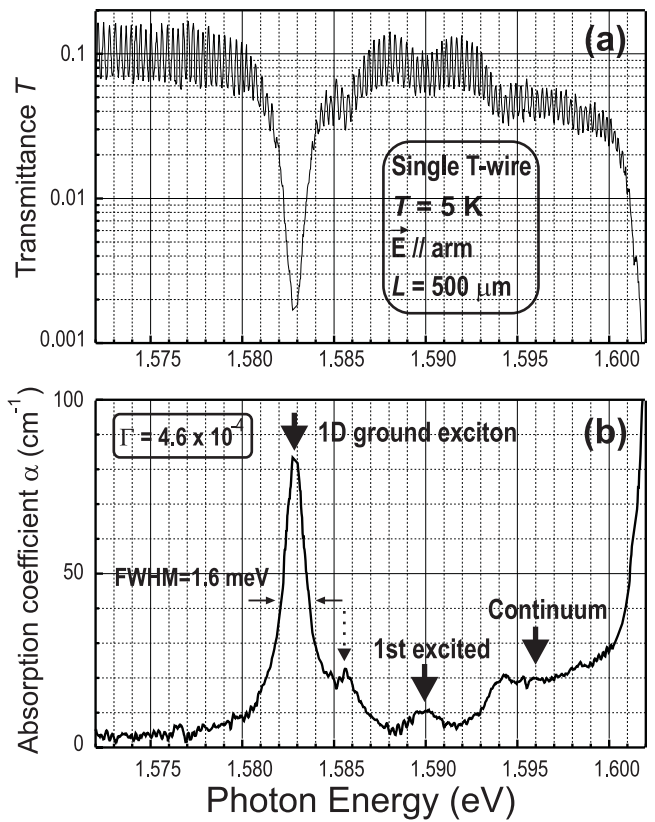}}}
\caption{(a) The transmittance spectrum $T(\lambda)$ of a single T-wire for light polarized parallel to the arm well at 5 K. (b) The absorption spectrum derived from the transmittance spectrum.}
\label{fig2}
\end{figure}

Figure \ref{fig2}(a) shows the transmittance spectrum $T(\lambda)$ for the single T-wire around the band gap at 5 K. We set $I_0(\lambda)$ to 200 nW or less to avoid absorption saturation, which occurs above 440 nW. The spectral resolution is 0.025 meV, as defined by the scanning steps of the laser wavelength. Strong attenuation is observed at 1.5828 eV and above 1.600 eV due to excitons in the T-wire and arm well, respectively. This will be explained later. In the other region, we observe clear Fabry-Perot oscillations.

The equation for the transmittance spectrum is given by
\begin{equation}
{T(\lambda)}=\frac{(1-R)^2 {\rm{e}}^{-\alpha (\lambda)L} \eta}{\{1-{\rm{e}}^{-\alpha (\lambda)L}R\}^2+4{\rm{e}}^{-\alpha (\lambda)L}R\sin^2\delta (\lambda)}\ \ 
\label{etalon}
\end{equation}
where $R$ is reflectance of 0.295 corresponding to $n$=3.38, $L$ is the cavity length of 500 $\mu$m, $\alpha(\lambda)$ is modal absorption coefficient, $\eta$ is the coupling efficiency of the incident light to the T-waveguide, and $\delta(\lambda)\ (=2\pi nL/\lambda)$ is the round-trip phase difference in the cavity, respectively. As will be proven later, analysis of the Fabry-Perot fringes enables estimation of the coupling efficiency $\eta$, and the obtained value is $\eta$=0.24$\pm$0.05. The second term in the denominator can be ignored in large absorption coefficients.

In Eq. (\ref{etalon}), $\sin^2\delta(\lambda)$ can be replaced with 1/2 if we take the inversed average of $T(\lambda)$ over one oscillation period. Then, $\alpha(\lambda)$ can be derived from the equation without oscillation. Indeed, we took the inversed average of $T(\lambda)$ in Fig. \ref{fig2}(a) over an oscillation period of 0.285 meV except for the strong-absorption regions without fringes, and derived $\alpha(\lambda)$ using $\eta$=0.24. 

Figure \ref{fig2}(b) shows the absorption spectrum for the single T-wire at 5 K. It should be noted that the line shape of the absorption spectrum agrees well with that of the PLE spectrum\cite{itoh2003}, and we are able to observe the assigned structures of the 1D-exciton ground state, the excited state of the 1D-exciton, and the 1D continuum states, as denoted in the figure. The absorption above 1.600 eV is due to the arm well. The absorption peak of 1D-exciton ground state has a full width at half maximum (FWHM) of 1.6 meV. The Stokes shift, that is the peak shift energy of PL and absorption peaks, is below 0.5 meV\cite{akiyama2003}. These demonstrate the high uniformity of our quantum wire in the whole region. It is remarkable that both sides of the exciton peak drop sharply to the background level. The small additional peak at around 1.5856 eV (dashed arrow) is due to the T-wire that consists of a 1-monolayer-thinner arm well\cite{yoshita2002}.

The absolute values of the absorption coefficient shown in Fig. \ref{fig2}(b) are the key results of this experiment. In the low-energy region below the 1D-exciton ground state, there is a background absorption of 3.5 cm$^{-1}$. This is likely due to waveguide loss and it adds a flat background to the whole spectrum. The absorption maximum for the 1D-exciton ground state shows $\alpha$=80 cm$^{-1}$, when we subtract the waveguide loss. The area of absorption coefficient for the peak is evaluated to be 190 cm$^{-1}$meV. The value of $\alpha$=80 cm$^{-1}$ gives e$^{-\alpha L}$=0.018, which means that 98 \% of the transmission light is absorbed by a single pass of the 500-$\mu$m-long waveguide. This demonstrates a strong excitonic absorption of the single T-wire, in spite of the small optical confinement factor $\Gamma=4.6\times 10^{-4}$. On the other hand, the absorption coefficient of 1D continuum states is about 16 cm$^{-1}$. This is much smaller than that of the 1D-exciton ground state, and the inverse-square-root singularity at the band edge is absent, as predicted theoretically\cite{ogawa1991}.

For a 7$\times$7 nm lateral-sized T-wire, Wang and Das Sarma calculated the material absorption coefficient $\alpha/\Gamma$ of the 1D-exciton ground state at 10 K\cite{wang2001}. From the figure shown in their paper, the area of material absorption coefficient for the peak is estimated to be $6.2\times10^2$ cm$^{-1}$eV. If we embed a 7$\times7$ nm T-wire into our T-waveguide, where $\Gamma$ is $2.70\times10^{-4}$, the area of the modal absorption coefficient becomes 167 cm$^{-1}$meV. This shows good agreement with our experimental result.

\begin{figure}[tbp]
\centerline{\scalebox{1.0}{\includegraphics{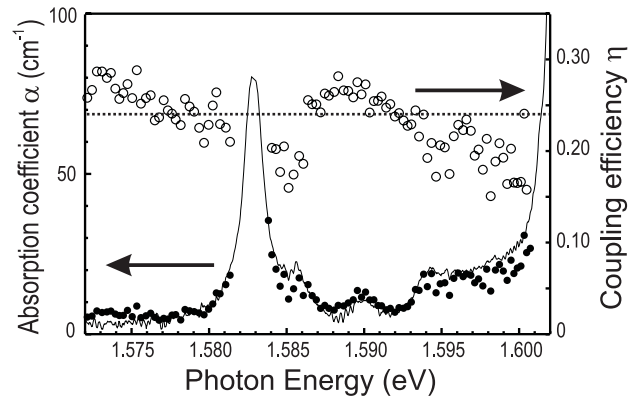}}}
\caption{A comparison of the absorption coefficient $\alpha$ derived using Hakki and Paoli's method (dots) with that shown in Fig. \ref{fig2}(b) (solid line). They correspond to the left longitudinal axis. Open circles with right longitudinal axis show the coupling efficiency $\eta$ derived using Hakki and Paoli's method. The dotted horizontal line represents $\eta$=0.24.}
\label{fig3}
\end{figure}

We verify the accuracy of $\alpha$ and $\eta$, based on the Fabry-Perot-fringe analysis method by Hakki and Paoli\cite{hakki1974}. In this analysis, $\alpha$ is derived from the ratio of transmittance maximum $T_{\rm{max}}$ and minimum $T_{\rm{min}}$ in each oscillation using $T_{\rm{max}}/T_{\rm{min}}=\{(1+{\rm{e}}^{-\alpha L}R)/(1-{\rm{e}}^{-\alpha L}R)\}^2$\cite{hakki1974}, and $\eta$ can then be derived from $T_{\rm{max}}$ or $T_{\rm{min}}$ in Eq. (\ref{etalon}). The analysis tends to overestimate $\alpha$ if the spectral resolution is not high enough and to have large scattering or error in the region of strong absorption \cite{jordan1994}, but it is useful for checking the internal consistency. In Fig. \ref{fig3}, the open circles ($\circ$) and dots ($\bullet$) plot $\eta$ and $\alpha$, respectively, using this analysis. This allow us to achieve a coupling efficiency of $\eta$=0.24$\pm$0.05. In the figure, the absorption spectrum of Fig. \ref{fig2}(b) derived for the constant $\eta$=0.24 is shown using a solid curve. The two absorption shapes are similar and the difference or experimental error is at most 3 cm$^{-1}$.

In summary, we measured the absolute values of the absorption coefficients for single-quantum-wire device using waveguide-transmission spectroscopy. The single quantum wire has a sharp, strong absorption peak at the energy of the 1D-exciton ground state for the light propagating along the wire, in spite of its small volume. This shows great promise for future applications of quantum wires to optical devices. 

This work was partly supported by a Grant-in-Aid from the Ministry of Education, Culture, Sports, Science, and Technology, Japan.


\end{document}